\documentclass[conference]{IEEEtran}
\IEEEoverridecommandlockouts

\usepackage{cite}
\usepackage{amsmath,amssymb,amsfonts,bbm}
\usepackage{graphicx}
\usepackage{textcomp}
\usepackage{xcolor}
\def\BibTeX{{\rm B\kern-.05em{\sc i\kern-.025em b}\kern-.08em
    T\kern-.1667em\lower.7ex\hbox{E}\kern-.125emX}}

\usepackage{amsthm}
\theoremstyle{definition} 

\theoremstyle{theorem}

\newtheorem{theorem}{Theorem}
\newtheorem{lemma}{Lemma}
\newtheorem{proposition}{Proposition}

\usepackage{caption}
\usepackage{booktabs}
\usepackage{multirow}
\usepackage{rotating}

\usepackage{tabularx}

\usepackage{mathtools}

\newcommand{\myitem}[1]{\vspace{2pt}\noindent\textbf{#1}}

\usepackage[colorlinks=true, allcolors=blue]{hyperref}

\usepackage[noend]{algorithmic}
\usepackage[linesnumbered,ruled,vlined,algo2e]{algorithm2e} %
\usepackage{algorithm}
\let\oldnl\nl
\newcommand{\nonl}{\renewcommand{\nl}{\let\nl\oldnl}}

\usepackage{color}
\definecolor{red}{rgb}{1,0.2,0.2}
\definecolor{green}{rgb}{0.2,1,0.5}
\definecolor{blue}{rgb}{0,0,1}
\definecolor{lightblue}{rgb}{0.3,0.5,1}

\newcommand{\Prob}[1]{\mathbb{P}(#1)}

\newcommand{\ket}[1]{|#1\rangle}
\newcommand{\bra}[1]{\langle#1|}

\title{Optimal Holding Times for Entanglement Swapping\\%
\thanks{Any mention of commercial products in this paper is for information only; it does not imply any recommendation or endorsement by NIST. U.S. Government work not protected by U.S. copyright.}
}

\author{
\IEEEauthorblockN{
Van Sy Mai\textsuperscript{1},
Richard J. La\textsuperscript{2},
Abdella Battou\textsuperscript{1}
}

\IEEEauthorblockA{
\textsuperscript{1}Communications Technology Laboratory, NIST, Gaithersburg, MD, USA,\\
\textsuperscript{2}Electrical and Computer Engineering, University of Maryland, College Park, MD, USA
}

\IEEEauthorblockA{
vansy.mai@nist.gov,
hyongla@umd.edu,
abdella.battou@nist.gov
}
}

\begin{document}
\maketitle

\begin{abstract}
We revisit the optimization of entanglement holding times in memory-based quantum repeaters for maximizing entanglement swapping throughput under different fidelity constraints -- strict, probabilistic, and mean. We consider two heterogeneous quantum links that are equipped with single or multiple memories and differ in decoherence rates, memory capacities, entanglement attempt rates and success probabilities, as well as classical communication latencies. Efficient models are developed to estimate the expected entanglement throughput and the probability density functions of the conditional wait time to swap. The models are then  used to satisfy different fidelity requirements by optimizing the maximum holding times for the two links. Finally, our  proposed model and approach are validated with a discrete-event quantum network simulator.
\end{abstract}

\section{Introduction}
Quantum networks are expected to support applications beyond the capabilities of classical networks, including blind quantum computing, distributed quantum sensing, quantum key distribution, distributed quantum computation, and ultimately the quantum internet \cite{kimble2008quantum, wehner2018quantum}. These applications require the distribution of high-quality end-to-end (E2E) entanglement over long distances, typically through networks of quantum memories. Achieving this remains challenging because entangled states are vulnerable to loss, environmental decoherence, and interface imperfections. Moreover, current quantum memories for communications have limited efficiency and coherence time, constraining both the distance and fidelity of E2E entanglement distribution.

Many applications impose certain fidelity requirements on the E2E entanglements they consume. We consider a simple network comprising two quantum links connected by a quantum repeater that performs entanglement swapping. Our objective is to develop a framework for maximizing the E2E entanglement distribution rate subject to strict, probabilistic, or mean fidelity requirements.

\myitem{Related work}: 
Prior work on homogeneous quantum-repeater chains characterizes the wait-time distribution for E2E entanglement \cite{shchukin2019waiting}. 
Polynomial-time methods in \cite{brand2020efficient,li2021efficient} additionally compute rate and fidelity distributions under Werner-state models; truncated distributions are then used to optimize repeater cutoff times and secret-key rates. For homogeneous chains with unlimited storage and communication qubits, \cite{de2024analysis} derives the average fidelity for sequential swapping. These assumptions ignore the effects of finite buffers, multiplexing, and heterogeneous cutoff times. Other studies consider different swapping strategies such as  
doubling \cite{caleffi2017optimal}, 
swap-asap \cite{kamin2023exact}, 
and heuristic policies \cite{mai2025towards}. 
Most use discrete-time models with homogeneous links and memories and represent fidelity degradation primarily through repeater cutoff times.

Queueing models have also been applied to quantum switches. The continuous-time Markov chain (CTMC) in \cite{vardoyan2019stochastic} characterizes  switching capacity of a star network with equal buffer sizes, constant link-generation rates, and decoherence cutoffs. However, it does not optimize cutoff times under an explicit entanglement-noise model. Related matching-queue models consider general entanglement-arrival distributions, switch stability, and max-weight scheduling \cite{zubeldia2026matching,fittipaldi2023linear}. 
The work \cite{perry1999perishable} studies finite inventories of perishable items with state-dependent arrivals and deterministic or exponential patience. Recently, \cite{mai2026swapping} proposed a throughput model using double-ended queues with renewal arrivals and abandonment under a strict fidelity constraint but does not study different matching policies. 



\myitem{Contributions}: 
We extend the work in \cite{mai2026swapping} by developing a new fidelity model and considering more practical fidelity constraints. Specifically, we propose an analytical model to estimate the probability density function (PDF) of E2E entanglement fidelity in multi-memory queues with different matching policies subject to maximum holding times (MHTs) of the stored entangled qubits at the repeater. Second, using the above models, we formulate the problem of optimizing the MHTs to maximize the E2E entanglement throughput subject to strict, probabilistic or mean constraints on E2E entanglement fidelity and derive optimal solutions. Finally, we present simulation results obtained using a discrete event quantum network simulator to validate our models and insights.

The rest of this paper is as follows. Section II provides network and noise assumptions and problem formulation. Section III details queueing models for computing the expected entanglement throughput and PDF of conditional wait time to swap. Section IV presents our solution to maximizing throughput subject to fidelity constraints. Section V illustrates its efficacy through simulation via the quantum network simulator SeQUeNCe, followed by conclusions in Section VI.

\myitem{Notation:} $\mathbb{E}[X]$ denotes the expected value of a random variable (RV) $X$. 
For any $x\in \mathbb{R}$, $[x]_+ := \max(0,x)$. 

\section{Link Model and Problem Formulation}
\label{sec:Network_Model}

We consider a three-node quantum path $0$--$1$--$2$ in which each node has a finite number of memories. Adjacent nodes generate elementary Bell pairs over lossy quantum channels, and node~$1$ swaps pairs on links $(0,1)$ and $(1,2)$, also referred to as links 0 and 1 for short, to create E2E entanglement. Classical connectivity is available for heralding and control. We focus on entanglement generation and swapping and do not consider purification.

\subsection{Assumptions and Setup}

{\bf A1.} On each link, when end-node memories are available, (entanglement) pairs are generated via an independent Poisson process with a potentially state-dependent rate.

{\bf A2.} Due to memory decoherence, we limit the holding time of elementary entanglements on link $i$ to $W_i$ to ensure fidelity.

{\bf A3.}  Upon matching elementary (entanglement) pairs on two links, swapping occurs immediately. This process takes $T_{\rm swp}$ and succeeds with probability $q \in (0, 1]$. Outcomes are heralded to nodes 0 and 2 with delays $\ell_{01}$ and $\ell_{12}$, respectively. We assume that nodes consume qubits immediately upon heralding, freeing up memories for new generation attempts.

{\bf A4.} Link attempts are generally initiated by a fixed node. For instance, if node 1 starts link 0, a new attempt can piggyback on heralding messages for an immediate start. However, if node 0 initiates, a delay of $\ell_{01}$ is incurred. We denote this potential reset delay after swapping as $T_i$ for link~$i$.

We describe each link $i=0,1$ using $$\mathcal{Q}_i = (K_i, {\lambda}_i, W_i, T_i),$$ 
where $K_i$ is the memory capacity, and ${\lambda}_i \in \mathbb{R}^{K_i}$ is the vector of entanglement arrival rates at different states. 


\subsection{Noise and Fidelity Models}
We denote 1-qubit depolarizing and dephasing channels by 
\begin{align*}
    \mathcal{E}_{v}^{(1)}(\rho) = v \rho + (1{-}v)\frac{I_2}{2},\quad
    \mathcal{F}_{w}^{(1)}(\rho) = \frac{1 {+} w}{2}\rho + \frac{1 {-} w}{2} \sigma_z\rho \sigma_z
\end{align*}
for $v,w\in[0,1]$, where $\rho$ is a qubit state, and $\sigma_z$ the Pauli-Z matrix. These channels commute, and parameters multiply under composition. For independently affected qubits, the corresponding 2-qubit channels are tensor products; when applied to a Bell state, their effects depend only on the products of the two 1-qubit parameters. 
Specifically, the Bell state $\rho_{\Psi^+}=\ket{\Psi^+}\!\bra{\Psi^+}$ under depolarizing and dephasing channels are respectively described by
\begin{align*}
    \mathcal{E}_{v_1v_2}^{(2)}(\rho_{\Psi^+}) &=v_1v_2\rho_{\Psi^+}+(1-v_1v_2)\frac{I_4}{4}\\
    \mathcal{F}_{w_1w_2}^{(2)}(\rho_{\Psi^+}) &=\frac{1+w_1w_2}{2}\rho_{\Psi^+} +\frac{1-w_1w_2}{2}\rho_{\Psi^-}.
\end{align*}

Memory decoherence at node $i$ over time $t$ is given by the composition $\mathcal{D}_i(t)=\mathcal{E}_{v_i(t)}^{(1)}\circ \mathcal{F}_{w_i(t)}^{(1)}$, where $v_i(t)$ and $w_i(t)$ are nonincreasing. This includes, for example, exponential decay under Markovian noise and Gaussian dephasing under slow fluctuations. A successful heralded entanglement-generation signals an elementary pair on link $(i,j)$ of age $t$ with state
\begin{align*}
    \rho_{ij}(t)
    =\bigl(\mathcal{D}_i(t)\otimes\mathcal{D}_j(t)\bigr)
    \circ\bigl(\mathcal{E}_{v_{ij}^{\rm B}}^{(2)}
    \circ\mathcal{F}_{w_{ij}^{\rm B}}^{(2)}\bigr)
    (\rho_{\Psi^+}),
\end{align*}
where $v_{ij}^{\rm B}$ and $w_{ij}^{\rm B}$ capture link and measurement errors. 
Local correction errors are absorbed into these effective parameters.

When both elementary links are available, node~$1$ performs a swap. The operation takes $T_{\rm swp}$ time, may be probabilistic, and contributes a depolarizing factor $v_1^{\rm swp}$. Let $a_{1L}$ and $a_{1R}$ denote the ages of the two memories at node~$1$ when they are swapped. The effective post-swap parameters are $v_{02}^{\rm B}=v_{01}^{\rm B}v_{12}^{\rm B} v_1(a_{1L})v_1(a_{1R})v_1^{\rm swp}$ for depolarization and $w_{02}^{\rm B}=w_{01}^{\rm B}w_{12}^{\rm B} w_1(a_{1L})w_1(a_{1R})$ for dephasing. Accounting for end-memory ages $(a_{0R}, a_{2L})$ at heralding, 
let
\begin{align*}
    v_{02}=v_0(a_{0R})v_2(a_{2L})v_{02}^{\rm B},\quad 
    w_{02}=w_0(a_{0R})w_2(a_{2L})w_{02}^{\rm B}.
\end{align*}
The resulting E2E state and fidelity are given by (see \cite{mai2026swapping})
\begin{align}
\rho_{02} 
    &= v_{02}\Big(
      \frac{1+w_{02}}{2}\rho_{\Psi^+}
      +\frac{1-w_{02}}{2}\rho_{\Psi^-}
      \Big)
      +(1-v_{02})\frac{I_4}{4}, \nonumber\\
    f_{e2e}
    &=\bra{\Psi^+}\rho_{02}\ket{\Psi^+}
     =\left(1+v_{02}(1+2w_{02})\right)/4.
    \label{e2e_fidelity}
\end{align}

\subsection{Problem Formulation}
Under the above noise model, the E2E fidelity decreases with wait times for swapping entangled pairs; longer wait causes lower fidelity. We aim to maximize the exptected entanglement throughput by optimizing quantum memory MHTs $(W_0, W_1)$ to meet different fidelity requirements:
\begin{align}
    \mbox{(strict)} \hspace{0.4in} &f_{e2e}\geq F_{\rm req}  \label{eq_Fmin}\\
    \mbox{(probabilistic)} \hspace{0.4in} &\Prob{f_{e2e}\geq F_{\rm desired}}\geq 1-\varepsilon \label{eq_Fdesired}\\
    \mbox{(mean)} \hspace{0.4in} &\mathbb{E}[f_{e2e}]\geq F_{\rm mean}. \label{eq_Fmean}
\end{align}

\section{Entanglement Rate and Wait Time Models}
    \label{sec_throughput}
We start with the simpler cases with a unit memory for each link, followed by more general cases with multiple memories. To simplify notation, for $i = 0, 1$, we use $\bar{i}$ to denote $1\!-\!i$. 

\subsection{Unit-Memory Case}
It has been shown \cite{mai2026swapping} that the unit-memory case can be analyzed efficiently and accurately using renewal reward theory~\cite{Ross}. Specifically, let 
$$T_p = T_{\rm swp}+\max(T_0, T_1), \quad \alpha_i = 1-e^{-\lambda_{\bar i}W_i}$$
Here $\alpha_i$ is the probability that, conditional on the event that an entanglement arrives at $\mathcal{Q}_i$ first, a match from $\mathcal{Q}_{\bar{i}}$ arrives within $W_i$. 
We have the following result \cite{mai2026swapping}.

\begin{theorem} \label{thm_unit_memory_rate}
For unit-memory queues, the expected rate is 
\begin{equation}
    R_{\rm um} = \frac{q(\alpha_0 \lambda_0 + \alpha_1 \lambda_1)}{1+\sum_{i=0}^{1}\alpha_i\lambda_i(T_p+\lambda_{\bar{i}}^{-1}) + (1-\alpha_i)\lambda_iT_i}.
    \label{eq_Rate_UnitCapacity}
\end{equation}
Let $\delta$ denote the conditional wait time to swap. Its PDF is   
\begin{equation}\label{eq_waittime_distr_unit}    
    g_\delta(x) 
    = p_{0|\rm m}\cdot g_{\mathtt{TE}}(x;\lambda_{1}, W_{0}) 
    + p_{1|\rm m}\cdot g_{\mathtt{TE}}(x;\lambda_{0}, W_{1}), 
\end{equation}
where $g_{\mathtt{TE}}(x;\lambda_{\bar{i}}, W_i) = \alpha_i^{-1} {\lambda_{\bar{i}} e^{-\lambda_{\bar{i}} x}} \mathbf{1}_{(0, W_i)}(x)$ 
and $p_{i|\rm m} = {\lambda_{i}\alpha_{i}}/{(\lambda_{0}\alpha_{0} + \lambda_{1}\alpha_{1})}$ for $i=0,1$. 
\end{theorem}
Basically, the rate is the average reward per cycle length and the wait time is a mixture of two truncated exponentials. 

\subsection{Multiple-Memory Queues}\label{subsec_CTMC}
With multiple memories for a link, entanglement generation can continue as long as there are available memories. The two links form a finite-capacity double-ended queue: an arrival is immediately matched when the opposite queue is nonempty and is otherwise stored until it is matched or expires. We consider three common matching policies First-In-First-Out, (FIFO), Last-In-First-Out (LIFO), and Random Selection (RAND), which affect pair ages and hence E2E fidelity. 

\subsubsection{Estimating Throughput}
An effective 4D-CTMC model can be constructed by approximating memory holding times, swapping times, and reset delays as independent exponential RVs \cite{mai2026swapping}. Its state is $(n,m_p,m_0,m_1),$ where $n$ is the (signed) net number of waiting pairs, $m_p$ the number of pairs undergoing swapping, and $m_i$ the number of memories being reset on link $i$. However, this might become less efficient as the number of memories grows; e.g., the system with $K_i\equiv K$ has $O(K^4)$ states. Moreover, the full state becomes unnecessary when swapping and reset delays are short relative to link-generation times. Thus, we use a one-dimensional Birth-Death Process (BDP) for throughput analysis by setting $m_p=m_0=m_1=0$ and retaining only $n\in\{-K_0,\ldots,K_1\}.$  This is reasonable in the multiple-memory regime because, unlike the unit-memory case, a reset does not normally create a queue-wide idle period; only the affected memory is temporarily unavailable while other free memories can continue new generation attempts. 

Let $\theta_i=1/W_i$ be the exponential expiration rate. The BDP stationary probabilities are given as follows  (see e.g., \cite{perry1999perishable,mai2026swapping})
\begin{align}
    \pi_{-m} 
    =\pi_0\prod_{k=1}^{m} \frac{\lambda_{0|k-1}}{\lambda_{1|0}+k\theta_0},\quad 
    \pi_n
    =\pi_0\prod_{k=1}^{n} \frac{\lambda_{1|k-1}}{\lambda_{0|0}+k\theta_1}
    \label{eq_pi_n}
\end{align}
for $m=1,\ldots,K_0$ and $n=1,\ldots,K_1$, where 
\begin{equation}
    \pi_0={(1+E_0+E_1)^{-1}}, ~~ E_i=\sum_{n=1}^{K_i}\prod_{k=1}^{n}
    \frac{\lambda_{i|k-1}}{\lambda_{\bar{i}|0}+k\theta_i}. \label{eq_pi_0}
\end{equation}
\begin{theorem}(\!\!\cite{mai2026swapping})\label{thm_rate_multi_mem}
For multi-memory queues, the entanglement throughput is approximated by
\begin{equation}
    R_{\rm bdp} =q \frac{\lambda_{1|0}E_0+\lambda_{0|0}E_1 }{1+E_0+E_1}. \label{eq_Rate_BDP}
\end{equation}
If $\lambda_{i|0} =\max_{k\ge0} \lambda_{i|k}$ for $i=0,1$, then $R_{\rm bdp}$ is component wise increasing in $W_i$. 
\end{theorem}

\subsubsection{Estimating Conditional Wait Time} 
The BDP model above only describes the macro-dynamics of the system, providing the count; it does not give the age of the pair selected for matching or impose bounds on holding times. To better capture the effects of matching policy and deterministic MHTs, we model the wait time of a `tagged' pair in more detail on top of the count BDP model; to estimate policy-specific wait times, we track a tagged pair from its admission until it would be matched if it could wait indefinitely, and then compare its wait time with its MHT. Denote by $\tau_i^{\rho}$ the wait time of this tagged pair arriving on link $i$. It is matched if and only if (iff) $\tau_i^{\rho} \le W_i$, in which case the matched age $\delta_i$ is $\tau_i^\rho$.

\myitem{Tagged-item absorbing CTMC}: 
We use a CTMC to track the tagged pair at $\mathcal{Q}_i$ with its state given by 
\begin{equation}
        (u,k) \text{\ for FIFO/LIFO}
        \quad\text{and~} u  \text{\ for RAND}
\end{equation}
where $u$ is the number of $\mathcal{Q}_i$ waiting pairs including the tagged pair, and $0\le k < u \le K_i$. For FIFO (LIFO), $k$ is the number of older (younger) pairs ahead of the tagged pair. We assume non-tagged pairs at $\mathcal{Q}_i$ expire at the effective rate $\theta_i = 1/W_i$.

\begingroup
\setlength{\tabcolsep}{4pt} 
\begin{table}[!b]
    \centering
    \resizebox{0.99\columnwidth}{!}{%
    \begin{tabular}{c|l|l|l|c}
    \toprule
    pol.  &  event & cond. & next state & rate \\
    \midrule
    \multirow{5}{*}{\begin{turn}{90} LIFO \hfill \end{turn}} &
    $\mathcal{Q}_i$ arrival & $u{<}K_i$ & $(u{+}1,k{+}1)$ & $\lambda_{i|u}$\\
    &$\mathcal{Q}_{\bar i}$ arrival matches tagged pair & $k{=}0$ & \text{matched} & $\lambda_{\bar i|0}$\\
    &$\mathcal{Q}_{\bar i}$ arrival matches younger pair & $k{>}0 $& $(u{-}1,k{-}1)$ & $\lambda_{\bar i|0}$\\
    &\text{Younger pair expires} & $k{>}0$ & $(u{-}1,k{-}1)$ & $k\theta_i$\\
    &\text{Older pair expires} & $u{-}k{>}1$ & $(u{-}1,k) $& $(u{-}1{-}k)\theta_i$\\
    \midrule 
    \midrule
    \multirow{5}{*}{\begin{turn}{90} FIFO \hfill \end{turn}}
    &$\mathcal{Q}_i$ arrival & $u{<}K_i$ & $(u{+}1,k)$ & $\lambda_{i|u}$\\
    &$\mathcal{Q}_{\bar i}$ arrival matches tagged pair & $k{=}0$ & \text{matched} & $\lambda_{\bar i|0}$\\
    &$\mathcal{Q}_{\bar i}$ arrival matches older pair & $k{>}0 $& $(u{-}1,k{-}1)$ & $\lambda_{\bar i|0}$\\
    &\text{Older pair expires} & $k{>}0$ & $(u{-}1,k{-}1)$ & $k\theta_i$\\
    &\text{Younger pair expires} & $u{-}k{>}1$ & $(u{-}1,k) $& $(u{-}1{-}k)\theta_i$\\
    \midrule
    \midrule 
    \multirow{4}{*}{\begin{turn}{90} RAND \hfill \end{turn}} &
    $\mathcal{Q}_i$ arrival & $u{<}K_i$ & $u{+}1$ & $\lambda_{i|u}$\\
    &$\mathcal{Q}_{\bar i}$ arrival matches tagged pair & $u{>}0$ & \text{matched} & $\lambda_{\bar i|0}/u$\\
    &$\mathcal{Q}_{\bar i}$ arrival matches diff. pair & $u{>}1 $& $u{-}1$ & ${(u{-}1)}\lambda_{\bar i|0}/{u}$\\
    &\text{Another pair expires} & $u{>}1$ & $u{-}1$ & $(u{-}1)\theta_i$\\
    \bottomrule
    \end{tabular}%
    }
    \caption{\small State transitions for different matching policies.} 
    \label{table_tagCTMC_state_transition}
\end{table}
\begingroup


For a matching policy $\rho$, let $H_i^\rho$, $b_i^{\rho}$ and $\nu_i^{\rho}$ denote the sub-generator, matched absorption vector, and the initial distribution, respectively, of tagged pair entering $\mathcal{Q}_i$: $H_i^\rho$ can be found from the state transitions under policy $\rho$; see Table~\ref{table_tagCTMC_state_transition} for the transition rates under different matching policies. Note that there is an absorbing state, namely \textit{matched}. The absorption rate vector to the {\em matched} state is given by
$$
    b_i^{\rho} = -H_i^{\rho} \mathbf{1} = \begin{cases}
    \lambda_{\bar i|0}\mathbf 1_{\{k=0\}} & \rho \in \{\rm LIFO, FIFO\} \\
    (\lambda_{\bar i|0}/u) \mathbf{1} &\rho = {\rm RAND} 
    \end{cases}
$$
and the initial states for FIFO, LIFO, and RAND are 
$(u_{\rm pre}+1, u_{\rm pre}), (u_{\rm pre}+1, 0)$, and $(u_{\rm pre}+1)$, respectively. Here $u_{\rm pre}$ denotes the pre-arrival state just before the tagged pair arrives at $\mathcal{Q}_i$ and does not match immediately, which happens at $\mathcal{Q}_0$ and $\mathcal{Q}_1$, respectively, with probabilities
\begin{align*}
    \frac{\pi_{-u_{\rm pre}}\lambda_{0|u_{\rm pre}}}{s_0}\quad \text{and}\quad \frac{\pi_{u_{\rm pre}}\lambda_{1|u_{\rm pre}}}{s_1} 
\end{align*} 
where $\pi$ is the stationary distribution of the BDP in \eqref{eq_pi_n}-\eqref{eq_pi_0} and $s_i$ is the arrival rate without immediate matching, which is given by 
\begin{align*}    
    s_0 = \sum_{u_{\rm pre}=0}^{K_0-1}\pi_{-u_{\rm pre}}\lambda_{0|u_{\rm pre}}, \quad
    s_1=  \sum_{u_{\rm pre}=0}^{K_1-1} \pi_{u_{\rm pre}}\lambda_{1|u_{\rm pre}}.
\end{align*}

Using the above, we can compute the initial distribution $\nu_i^{\rho}$ of the tagged pair state at $\mathcal{Q}_i$ under policy $\rho$. The rate at which the tagged pair enters the {\em matched} absorbing state at time $t$ is $\xi_i^{\rho}(t) = \nu_i^{\rho} e^{H_i^{\rho} t}b_i^\rho$ and, hence, the probability of being matched by time $t$ is equal to 
$$
C_i^{\rho}(t) = \int_0^t \xi_i^{\rho}(s) \mathrm{d}s = \nu_i^{\rho} (-H_i^{\rho})^{-1}(I-e^{H_i^{\rho} t})b_i^\rho.$$ Using $b_i^{\rho} = -H_i^{\rho} \mathbf{1}$, $\nu_i^{\rho} \mathbf{1}=1$ and $e^{H_i^{\rho} t}H_i^{\rho} = H_i^{\rho} e^{H_i^{\rho} t}$, we obtain
\begin{align}\label{eq_prob_matching_by_t}
    \xi_i^{\rho}(t) = \nu_i^{\rho} e^{H_i^{\rho} t}b_i^\rho \quad\text{and}\quad C_i^{\rho}(t) =1 - \nu_i^{\rho} e^{H_i^{\rho} t} \mathbf{1}. 
\end{align}
Since the tagged pair matches iff $\tau_i^\rho\le W_i$, the conditional cumulative distribution function (CDF) of wait time of matched tagged pair at $\mathcal{Q}_i$ before $W_i$ is $F_i^\rho(t;W_i) = {C_i^\rho(t)}/{C_i^\rho(W_i)}$ and the conditional PDF of wait time is
\begin{equation}\label{eq_pdf_TAG}
    g_{\mathtt{TAG}}^{\rho}(x; \lambda_{\bar{i}|0}, \mathcal{Q}_i) 
    = \frac{\xi_i^{\rho} (x)  \mathbf{1}_{(0, W_i)}(x)}{C_i^\rho(W_i)}  . 
\end{equation}
Putting together, the PDF of wait time $\delta= \delta_0 + \delta_1$ is the following mixture that resembles \eqref{eq_waittime_distr_unit}: 
\begin{equation}\label{eq_waittime_distr_multi}
g_\delta(x) 
    = p_{0|\rm m}  g_{\mathtt{TAG}}^{\rho}(x; \lambda_{1|0}, \mathcal{Q}_0)
    + p_{1|\rm m} g_{\mathtt{TAG}}^{\rho}(x; \lambda_{0|0}, \mathcal{Q}_1) 
\end{equation}
with the conditional probability $p_{i|\mathrm{m}} = {R_{i\gets \bar{i}}}/{(R_{i\gets \bar{i}} + R_{\bar{i}\gets i})}$, 
$R_{1 \gets 0} = \lambda_{0|0}\sum_{n=1}^{K_1}   \pi_{n} = \pi_0 \lambda_{0|0} E_1$ and $R_{0 \gets 1} = \pi_0 \lambda_{1|0} E_0 $. 

\section{Maximizing Rate with Fidelity Constraints}
\label{sec_maximize_Rate_Fidelity}

To simplify the exposition, we consider Markovian dephasing memories with $w_i(t) = e^{-\gamma_i t}$. In this case, 
\begin{equation*}
    w_{02} =w_{01}^{\rm B}w_{12}^{\rm B} e^{-A} \text{~with~}A = \gamma_0a_{0R} + \gamma_2a_{2L} + \gamma_1(a_{1L}+a_{1R})
\end{equation*}
which can be viewed as the total age of all qubits weighted by their dephasing rates. Recall that $a_{1L}$ and $a_{1R}$ are the ages of swapped qubits at node~$1$ and $a_{0R} = a_{1L}{+}T_{\rm swp}{+}\ell_{01}$ and $a_{2L} = a_{1R}{+}T_{\rm swp}{+}\ell_{12}$. 
Thus, 
\[
    {A} =  \gamma_0(\ell_{01}+T_{\rm swp}) + \gamma_2(\ell_{12}+T_{\rm swp})  + \gamma_{01}a_{1L}+ \gamma_{12}a_{1R}, 
\]
where  $\gamma_{i ,j} = \gamma_i +\gamma_{j}$ denotes the combined dephasing rate of the pair on link $(i,j)$. The ages $a_{1L}$ and $a_{1R}$ of the swapped qubits depend on which entangled pair is first heralded at node~$1$, and how long it has to wait for the other to be heralded. In our queueing model, this means that they depend on which queue experiences an arrival first and how long it must wait for an opposite arrival. Recall that $\delta_{0}\le W_0$ and $\delta_{1}\le W_1$ denote the wait times of pairs in $\mathcal{Q}_0$ and $\mathcal{Q}_1$ prior to a match, respectively, with $\delta_{0}\cdot\delta_{1}=0$. Thus, we can write 
\begin{align}\label{eq_total_age_3node_path_RV}
    {A} &= {A}_0 + Z, \quad \text{with}~ Z= \gamma_{01}\delta_0 + \gamma_{12}\delta_1\\
    {A}_0 &= \gamma_0\ell_{01} + \gamma_2\ell_{12} + \gamma_{02}T_{\rm swp} + \gamma_{01}\tau_{01}+ \gamma_{12}\tau_{12} .  \nonumber
\end{align}
Here, ${A}_0$ is fixed while $Z$ is the random part of $A$. 
Clearly, $f_{\mathrm{e2e}}=c_1+c_2e^{-Z}$ with $c_1{=}\tfrac{1+v_{02}}{4}$ and $c_2{=}\tfrac12 v_{02}w^{\rm B}_{01}w^{\rm B}_{12}e^{-A_0}$. 
The inverse map connecting fidelity to $Z$ is 
\begin{equation}
    \Theta(F) = \ln (2 {v}_{02} w^{\rm B}_{01}w^{\rm B}_{12}) - \ln(4F -1-{v}_{0n}) -A_0.
\end{equation}
Let us now consider the fidelity constraints \eqref{eq_Fmin}--\eqref{eq_Fmean} separately. 

\subsection{Strict Fidelity Constraint}
First, the ideal case with $Z=0$ dictates the maximum E2E fidelity $F_{\max} = c_1+c_2=(1+v_{02}(1+2w_{01}^{\rm B}w_{12}^{\rm B} e^{-{A}_{0}}))/4$. Consider $F_{\rm req} < F_{\max}$. Let $Z_{\rm thr} = \Theta(F_{\rm req})$ and define
\begin{equation}\label{eq_W_threshold}
    \bar{W}_0 = Z_{\rm thr}/\gamma_{01}, \quad \bar{W}_1 = Z_{\rm thr}/\gamma_{12} . 
\end{equation}
It has been shown \cite{mai2026swapping} that 
$$
    (\bar{W}_0, \bar{W}_1) \in \mathrm{argmax}\,\{R(W_0, W_1)\,|\,W_i\le \bar{W}_i, f_{\rm e2e} \ge F_{\rm req} \}, 
$$
where $R(W_0, W_1)$ denotes the rate given by either \eqref{eq_Rate_UnitCapacity} or \eqref{eq_Rate_BDP}. Thus, MHTs $\bar{W}_i$ maximize the rate while ensuring $Z \le {Z}_{\rm thr}$ for any distribution or mixture of $Z$. 

\subsection{Probabilistic Fidelity Constraint}
We rewrite condition $\Prob{f_{e2e} \geq F_{\rm desired}} \ge 1-\varepsilon $ as an equivalent constraint on the total dephasing age $A$. Let $Z_{\rm thr} = \Theta(F_{\rm desired})$, and $\bar{W}_i$ are given by \eqref{eq_W_threshold}. Since $\delta_0 \cdot \delta_1 = 0$, 
\begin{align*}
    \Prob{Z \le {Z}_{\rm thr}|\mathrm{m}} = p_{0|\rm m} \Prob{\delta_{0} \le \bar{W}_0} 
    + p_{1|\rm m} \Prob{\delta_{1} \le \bar{W}_1}
\end{align*}
which can be computed efficiently using \eqref{eq_waittime_distr_unit} and \eqref{eq_waittime_distr_multi}. 

\subsubsection{Unit-Memory Cases}
First, $\Prob{\delta_i \le \bar{W}_i}$ is
\begin{align*}\label{eq_pdf_TE}
\int_{0}^{\bar{W}_i} g_{\mathtt{TE}}(x;\lambda_{\bar{i}}, W_i) \mathrm{d} x 
    = \begin{cases}
    \frac{1- e^{-\lambda_{\bar{i}} \bar{W}_i}}{1-e^{-\lambda_{\bar{i}} W_i} }     & \bar{W}_i < W_i\\
    1 & \bar{W}_i \ge W_i
    \end{cases}
\end{align*}
which increases with our optimization variables $W_i$.
The probability of violating the threshold is 
\[
\Prob{\delta_i > \bar{W}_i} = 1 - \Prob{\delta_i \le \bar{W}_i} = \frac{[ e^{-\lambda_{\bar{i}} \bar{W}_i} - e^{-\lambda_{\bar{i}} {W}_i}]_+}{(1-e^{-\lambda_{\bar{i}} W_i} )}. 
\]
Multiplying both sides by $\lambda_i\alpha_i$ with $\alpha_i =1-e^{-\lambda_{\bar{i}} W_i} $ yields 
$$
    \lambda_i\alpha_i\Prob{\delta_i > \bar{W}_i} = [ \lambda_i e^{-\lambda_{\bar{i}} \bar{W}_i} - \lambda_i e^{-\lambda_{\bar{i}} {W}_i}]_+ = [\lambda_i\alpha_i - U_i]_+
$$
with 
$U_i = \lambda_i (1-e^{-\lambda_{\bar{i}} \bar{W}_i})$. As a result, we can express the probabilistic fidelity constraint $\Prob{Z \le {Z}_{\rm thr}|\mathrm{m}} \ge 1-\varepsilon$ as
\begin{equation}
    [\lambda_0\alpha_0 - U_o]_+ + [\lambda_1\alpha_1 - U_1]_+ \le \varepsilon(\lambda_0\alpha_0 + \lambda_1\alpha_1)  . 
\end{equation}
Now rewrite the matching rate as $R_{\mathrm{m}} = \frac{\lambda_0\alpha_0 + \lambda_1\alpha_1}{d_0\lambda_0\alpha_0 + d_1\lambda_1\alpha_1 + d_2}$ with $d_0=T_p - \ell_{01} + \lambda_{1}^{-1} $, $d_1=T_p - \ell_{12} + \lambda_{0}^{-1}$, and $d_2=1+ \lambda_0\ell_{01} + \lambda_{1}\ell_{12}$. Let $x_i=\lambda_i\alpha_i$ be the new optimization variables strictly increasing in $W_i$. The problem then becomes
\begin{align} 
    \underset{0\le x_i \le \lambda_i}{\mathrm{max}}\quad & R(x) = \frac{x_0+x_1}{d_0 x_0 + d_1 x_1 +d_2} \label{eq_fidelity_eps} \\
    \mathrm{s.t.} \quad& [x_0 - U_o]_+ + [x_1 - U_1]_+ \le \varepsilon(x_0 + x_1). \nonumber
\end{align}
This linear-fractional program has a closed form solution. 
\begin{theorem}\label{thm_optimal_x}
    Without loss of generality, suppose $d_0 \le d_1$. Let $U_{\varepsilon} = (U_0 + U_1)/(1- \varepsilon)$. The optimal solution is 
\begin{equation} \label{eq_x_opt_stochastic_unit}
    x^* = \begin{cases}
        (U_{\varepsilon}-U_1, U_1) & \text{if}~ U_1+\lambda_0 \ge U_{\varepsilon}, \\
        (\lambda_0, U_{\varepsilon}-\lambda_0) & \text{if}~ U_1+\lambda_0 < U_{\varepsilon} \le \lambda_0+\lambda_1, \\
        (\lambda_0, \lambda_1) &\text{if}~ \lambda_0+\lambda_1 < U_{\varepsilon} . 
    \end{cases}
\end{equation}
\end{theorem}
\begin{proof}
First, note that under the assumption $d_0\le d_1$, $R$ is coordinate-wise strictly increasing in $x_1\ge0$ and $x_0\lambda_0$. This can be seen from its partial derivatives $\partial_{x_1} R = \frac{d_2+x_1(d_1-d_0)}{D}>0$ and $\partial_{x_0} R = \frac{d_2-x_0(d_1-d_0)}{D}$ with $D=(d_0 x_0 + d_1 x_1 +d_2)^2$. For $x_0 < \lambda_0$, we have $x_0(d_1-d_0)< \lambda_0(d_1-d_0) = 1+\lambda_0\ell_{01} - \lambda_0(\lambda_1^{-1}+ \ell_{12}) < d_2.$ Thus $\partial_{x_0} R>0$.

Second, we argue that any solution must satisfy $x_i \ge  U_i$. 
To see this, for any feasible $x_i <  U_i$, consider any $\tilde{x}_i \in (x_i, U_i]$, which improves the objective $R$ while still satisfying~\eqref{eq_fidelity_eps}. This means that any $x_i <  U_i$ is not optimal. Thus, we must have $x_i \ge  U_i$ for $i=0,1$, and under this, \eqref{eq_fidelity_eps} can be replaced by
\begin{equation}\label{eq_sum_x}
    x_0 + x_1  \le (U_0 + U_1)/(1- \varepsilon) = U_{\varepsilon}. 
\end{equation}
Because of the bound $x_i \le \lambda_i$, consider two cases. First, if $U_{\varepsilon} > \lambda_0+\lambda_1$, then \eqref{eq_sum_x} is never active. By monotonicity of $R$, the optimal solution is $ x_i^*= \lambda_i.$ Second, if $U_{\varepsilon} < \lambda_0{+}\lambda_1$, we argue that \eqref{eq_sum_x} must be active at an optimal point. This can be seen again by contradiction: if $x_0 + x_1  < U_{\varepsilon}$, then we can increase either $x_i$ to increase $R$ while remaining feasible, proving sub-optimality in this case. Thus, it must hold $x_0 + x_1  = U_{\varepsilon}$. On this manifold, maximizing the rate becomes $${\mathrm{min}}_{x_i}\{ (d_1-d_0) x_1\,|\, x_0 + x_1 = U_{\varepsilon}, U_i \le x_i < \lambda_i \}.$$ 
It can be verified that the optimal solution is $ x^* =(U_{\varepsilon}-U_1, U_1)$ if $U_1+\lambda_0 \ge U_{\varepsilon}$ and $x^* = (\lambda_0, U_{\varepsilon}-\lambda_0)$ otherwise. Combining these two cases yields \eqref{eq_x_opt_stochastic_unit}. 
\end{proof}

The condition $d_0\le d_1$ is the same as $\lambda_1^{-1}+\ell_{12} < \lambda_0^{-1}+\ell_{01}$, which roughly means that the average wait-time to the first arrival of $\mathcal{Q}_1$ is less than that of $\mathcal{Q}_0$, i.e., link 1 has higher generation rate than link 0. In this case, for small $\varepsilon$, we must have $W_1^* = \bar{W}_1$ and only let $W_0^*$  violate  $\bar{W}_0$, allowing the slower link to stay slightly longer to meet probabilistic fidelity guarantees.

\subsubsection{Multi-Memory Cases}
Let us first reformulate the problem with strict MHTs. 
Recall that $R_{i \gets \bar{i}}(W)= \pi_0 \lambda_{\bar{i}|0} E_i$ is the rate of $\mathcal{Q}_{\bar{i}}$ arrivals matching $\mathcal{Q}_i$ with $W = (W_0, W_1)$. The total matching rate is $R_{\rm m}(W) = R_{0 \gets 1}(W) + R_{1 \gets 0}(W)$, 
$$
R_{\rm m}(W) =\frac{ \lambda_{1|0}E_{0}(W_0)+ \lambda_{0|0} E_{1}(W_1) }{1+  E_{0}(W_0) + E_{1}(W_1)}.
$$
Next, recall that the conditional CDF of the wait time of a matched pair that was waiting at $\mathcal{Q}_i$ is given by $F_i^\rho(z;W_i) = {C_i^{\rho}(z)}/{C_i^{\rho}(W_i)}$, where $C_i^{\rho}$ is computed in \eqref{eq_prob_matching_by_t}. Then, the probability of exceeding $\bar{W}_i$ must satisfy
\begin{equation}\label{eq:phi_exact}
    \Phi^\rho(W) = w_0(W) S_0^\rho(W_0) + w_1(W) S_1^\rho(W_1) \le \varepsilon, 
\end{equation}
where $S_i^\rho(W_i)=1-F_i^\rho(\bar{W}_i;W_i)$ 
and $w_i(W) = \frac{ R_{i \gets \bar{i}}(W) 
}{R_{\rm m}(W)}$,i.e., 
$$
w_i(W) = \frac{\lambda_{1|0}E_{0}(W_0)}{\big(\lambda_{1|0}E_{0}(W_0)+ \lambda_{0|0} E_{1}(W_1)\big)}.
$$
Therefore, the problem can be formulated as follows:
\begin{equation}\label{prob_maxRate_chanceFidelity}
    \mathrm{max}_{\,W}\ \{R_{\rm m}(W)  \mid \Phi^{\rho}(W) \le \varepsilon \}. 
\end{equation}

Unlike previous cases, a closed-form solution is complex. 
However, this is a nonlinear 2D problem that can be solved effectively by numerical solver; see Appendix for details. Here, for small $\varepsilon$, we hypothesize that the optimal solution is a perturbation of the strict-constraint solution, $\bar{W}_i$. 

For small $\varepsilon>0$, let $W_i = \bar{W}_i + \Delta_i$ with small $\Delta_i\ge 0$. Since $C_i^{\rho}(\bar{W}_i + \Delta_i) = C_i^{\rho}(\bar{W}_i) + \xi_i^{\rho}(\bar{W}_i)\Delta_i + O(\Delta_i^2)$,  
$$
S_i^{\rho}(W_i) 
= \frac{C_i^{\rho}(\bar{W}_i + \Delta_i) - C_i^{\rho}(\bar{W}_i)}{C_i^{\rho}(\bar{W}_i + \Delta_i)} = \beta_i^{\rho}\Delta_i + O(\Delta_i^2),
$$
where 
$\beta_i^{\rho} = \partial_{W_i} S_i^{\rho}(\bar{W}_i) = \frac{\xi_i^{\rho}(\bar{W}_i)}{C_i^{\rho}(\bar{W}_i)}$.  Because $S_i^\rho(\bar{W}_i)=0$,
\[
    \partial_{W_k} \left[w_i(W)S_i^\rho(W_i)\right]_{W=\bar{W}}
    = \bar{w}_i \partial_{W_k} S_i^\rho(\bar{W}_i) 
    = \bar{w}_i \beta_i^{\rho}
\]
with $\bar{w}_i = {w}_i(\bar{W})$. Thus, the threshold violation probability is 
\begin{equation*}
    \Phi^\rho(\bar{W}+\Delta)
    = \bar{w}_0 \beta_0^\rho\Delta_0 + \bar{w}_1 \beta_1^\rho\Delta_1 +  O(\|\Delta\|^2).
\end{equation*}
Let $g_i{=}\partial_{W_i} R_{\rm m}(\bar{W})>0$ by Theorem~\ref{thm_rate_multi_mem}. Expand $R_{\rm m}$ near $\bar{W}$:
\[
    R_{\rm m}(\bar{W} + \Delta) = R_{\rm m}(\bar{W})+g_0\Delta_0 +g_1\Delta_1 +O(\|\Delta\|^2).
\]
The first-order small-$\varepsilon$ optimization problem is given by 
\begin{equation}
    \mathrm{max}_{\Delta_i \ge 0}\ \{ g_0\Delta_0+g_1\Delta_1~|~ \bar{w}_0\beta_0^\rho\Delta_0
    + \bar{w}_1 \beta_1^\rho\Delta_1
    \le \varepsilon \}
\end{equation}
whose solution has a similar structure as the unit-memory case, i.e.,   
$
\Delta^* = \big(\frac{\varepsilon}{\bar{w}_0\beta_0^\rho}, 0\big) ~\text{if}~\frac{g_0}{\bar{w}_0\beta_0^\rho} > \frac{g_1}{\bar{w}_1\beta_1^\rho}~\text{and}~\big(0, \frac{\varepsilon}{\bar{w}_1\beta_1^\rho}\big)~\text{otherwise}. 
$

\subsection{Expected Fidelity Constraint}\label{subsec_Expected_F}
The multi-memory case is rather involved and we refer to Appendix~\ref{app:Multi-Memory} for discussion on nonlinear programming-based treatment.
Here, we focus on the unit-memory queues as it can be reformulated as a convex problem. Since $f_{\mathrm{e2e}}=c_1+c_2e^{-Z}$ with $c_1{=}\tfrac{1+v_{02}}{4}$ and $c_2{=}\tfrac12 v_{02}w^{B}_{01}w^{B}_{12}e^{-A_0}$, the constraint $\mathbb{E}[f_{\mathrm{e2e}}]\ge F_{\mathrm{mean}}$ is equivalent to 
\begin{equation}\label{eq_m_thr}
\mathbb{E}[e^{-Z}]\ge m_{\mathrm{thr}}:={(F_{\mathrm{mean}}-c_1)}/{c_2}.
\end{equation}

From \eqref{eq_total_age_3node_path_RV}, we have $Z = \sum_i \gamma^{(i)}\,\delta_i$ with $\gamma^{(0)}\!=\!\gamma_{01}$ and $\gamma^{(1)}\!=\!\gamma_{12}$. 
By \eqref{eq_waittime_distr_unit} and conditioning on which component is selected, 
\[
\mathbb{E}[e^{-Z}]
= \textstyle \sum_{i=0}^{1} p_{i|\mathrm{m}}\;
\mathbb{E}[e^{-\gamma^{(i)}\delta_i}].
\]
where $\mathbb{E} [e^{-\gamma^{(i)}\delta_i}] = \int e^{-\gamma^{(i)} x} g_{\mathtt{TE}}(x;\lambda_{\bar{i}}, W_i) \mathrm{d}x$, 
which is 
\[ 
\frac{\lambda_{\bar{i}}}{\alpha_i}
\int_0^{W_i} e^{-(\gamma^{(i)}+\lambda_{\bar{i}})x}\,\mathrm{d}x
= \frac{\lambda_{\bar{i}} \big(1-e^{-(\gamma^{(i)}+\lambda_{\bar{i}})W_i}\big)}
{\alpha_i\,(\gamma^{(i)}+\lambda_{\bar{i}})}.
\]
Multiplying by $p_{i|\mathrm{m}}$ and summing over $i$ yields
\[
    \mathbb{E}[e^{-Z}]
    =\frac{\lambda_0\lambda_1}{\lambda_0\alpha_0+\lambda_1\alpha_1}  \sum_{i} \frac{1-e^{-(\lambda_{\bar\imath}+\gamma^{(i)})W_i}}{\lambda_{\bar\imath}+\gamma^{(i)}}.
\]
As a result, \eqref{eq_m_thr} is the same as 
\begin{equation}\label{eq_EeZ}
 \sum_{i} \frac{\lambda_0\lambda_1(1-e^{-(\lambda_{\bar\imath}+\gamma^{(i)})W_i})}{\lambda_{\bar\imath}+\gamma^{(i)}} \ge (\lambda_0\alpha_0+\lambda_1\alpha_1)m_{\rm thr}.
\end{equation}
Let $s_i=1 - \alpha_i = e^{-\lambda_{\bar\imath}W_i}$ and $\kappa_i = 1 + \gamma^{(i)}\lambda_{\bar\imath}^{-1}$, and define 
$$
g_i(s_i)=m_{\mathrm{thr}}\lambda_i(1-s_i)-\frac{\lambda_0\lambda_1}{\lambda_{\bar\imath}+\gamma^{(i)}}(1-s_i^{\kappa_i}). 
$$
Then, \eqref{eq_EeZ} becomes 
$$\textstyle \sum_i g_i(s_i)\le 0, \quad $$ 
which is strictly convex since $g_i''(s_i){=}\lambda_i(\kappa_i-1)s_i^{\kappa_i-2}>0$. 

Because $R_{\rm um}$ is linear-fractional in $s$, by applying the transformation  
\[
\tau=1/(d_0\lambda_0\alpha_0+d_1\lambda_1\alpha_1 +d_2) \ \mbox{ and } 
\ y_i=\tau s_i
\]
and using epigraph variables $z_i\ge y_i^{\kappa_i}\tau^{1-\kappa_i}$, we can reformulate the optimization as the following exact convex optimization problem:
\begin{equation}
    \begin{aligned}
    \max_{\tau>0,y,z}\ &\ \textstyle \sum_i\lambda_i(\tau-y_i)\\
    \text{s.t.}\ &\ \textstyle\sum_i d_i\lambda_i(\tau-y_i)+d_2\tau=1 \\
    &\ \textstyle\sum_i\!\big[m_{\mathrm{thr}}\lambda_i(\tau-y_i)
       -\tfrac{\lambda_0\lambda_1}{\lambda_{\bar\imath}+\gamma^{(i)}}(\tau-z_i)\big]\le0 \\
    &\ (z_i,\tau,y_i)\in\mathcal{K}^{\zeta_i}_{\mathrm{pow}},\
       \zeta_i=\tfrac{\lambda_{\bar\imath}}{\lambda_{\bar\imath}+\gamma^{(i)}},\
       0\le y_i\le \tau 
    \end{aligned}
\end{equation}
where we recall $d_0 = T_p - \ell_{01} + \lambda_{1}^{-1} $, $d_1 = T_p - \ell_{12} + \lambda_{0}^{-1}$, and $d_2 = 1+ \lambda_0\ell_{01} + \lambda_{1}\ell_{12}$. 
Here,   
$$\mathcal{K}^{\zeta}_{\mathrm{pow}}=\{(a,b,c):a^{\zeta}b^{1-\zeta}\ge|c|,a,b\ge0\}$$ is the power cone. This problem can be solved by interior-point methods and we recover $W_i^\star=-\lambda_{\bar\imath}^{-1}\ln(y_i^\star/\tau^\star)$. Note that strict and probabilistic constraints in \eqref{eq_Fmin} and \eqref{eq_Fdesired}, respectively, can also be incorporated into this convex reformulation via suitable bounds on $W_i$ and thus $s_i$, which can then be translated into linear inequality in $y_i$ and $\tau$.

\section{Simulation and Evaluation}
\label{sec:evaluation}
\subsection{Setup and Parameters}
We evaluate our models and methods using a simulator SeQUeNCe \cite{wu2021sequence}; we focus on multi-memory with strict and probabilistic fidelity constraints. In all considered cases, we run $100$ independent simulations, each with a $10\,\mathrm{s}$ duration.

\begin{figure*}[!t]
    \centering
    \includegraphics[width=0.9\textwidth]{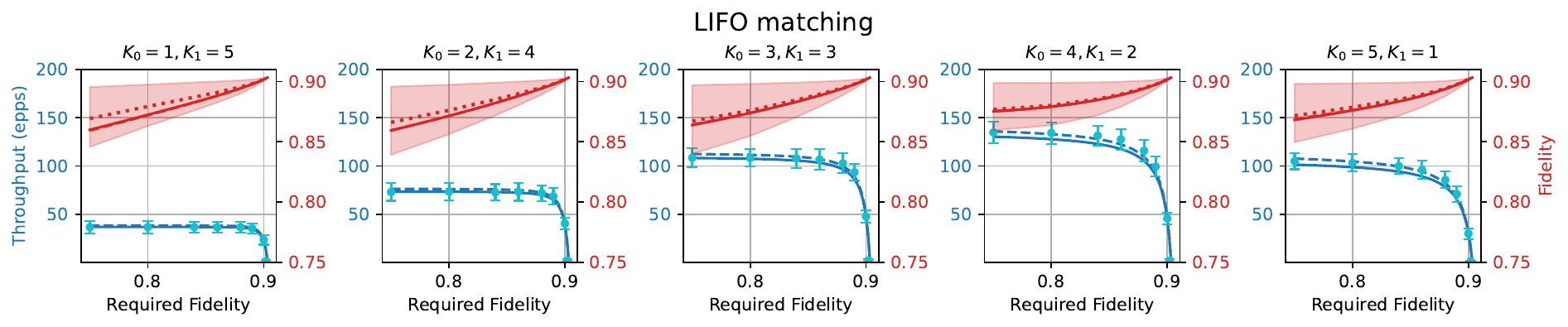}
    %
    \includegraphics[width=0.9\textwidth]{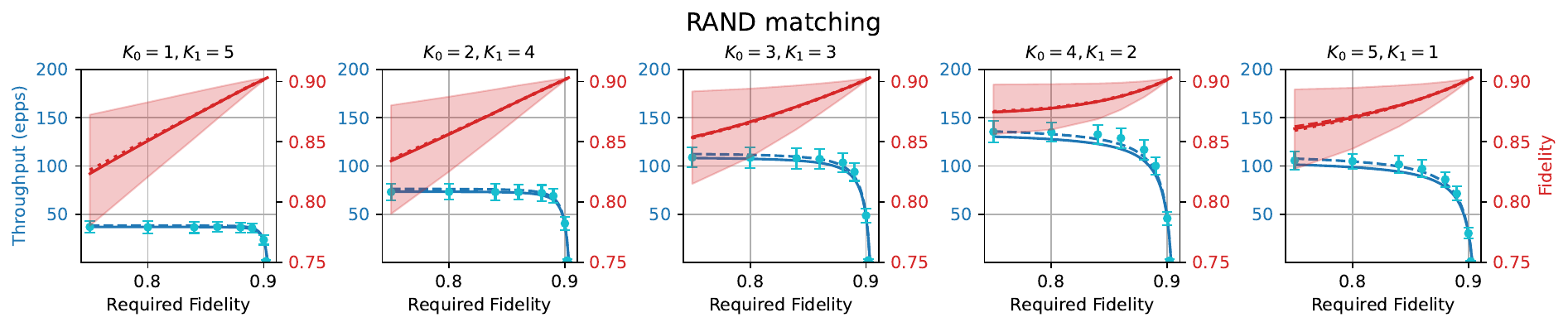}
%
    \includegraphics[width=0.9\textwidth]{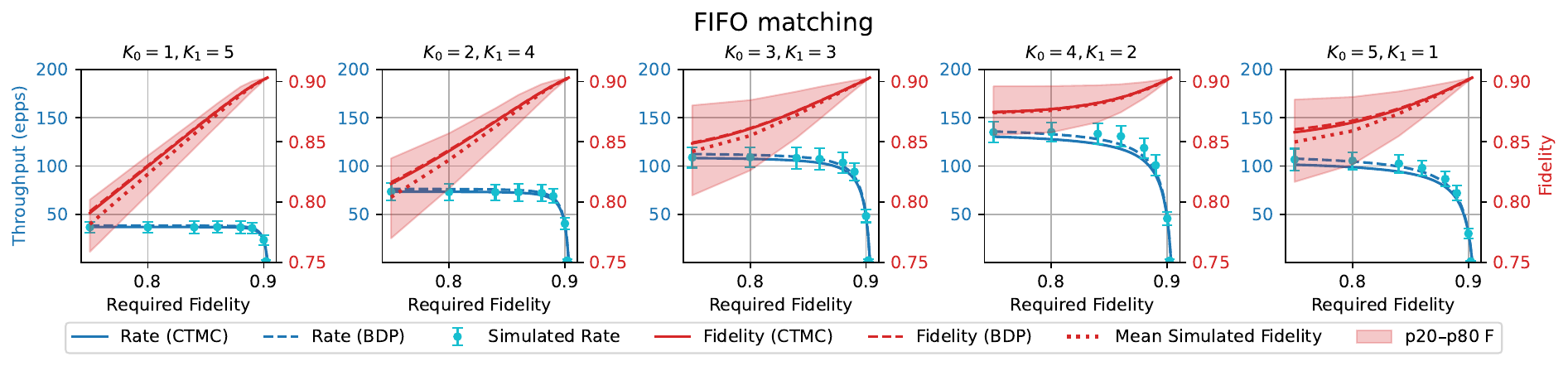}
    \caption{\small Expected rate and fidelity when varying minimum required fideltiy $F_{\rm req}$, memory allocation and matching policy.}
    \label{fig_multi_memory}
\end{figure*}

\textit{Channels and Bell State Measurements (BSMs)}: Quantum channels are optical fiber with  $L_{01}{=}32\,\text{km}$, $L_{12}{=}18\,\text{km}$, a peak frequency of $50$ MHz and attenuation rate of $0.2$ dB/km. The light speed in fiber is $c_0 = 2 {\times} 10^5\,\text{km/s}$. BSM dephasing parameters are modeled as $w_{ij}^{\rm B} = e^{-L_{ij}/2L_c}$ with a coherence length of $L_c=250\,\text{km}$. 
BSM depolarizing parameters due mainly to background photons are $v_{01}^{\rm B} = 0.03$ and $v_{12}^{\rm B} = 0.02$.   
We assume detectors at photonic BSMs with moderate parameters: $100\,\text{ps}$ resolution, $\eta_{\rm d}=0.85$ efficiency, and  $50\,\text{Hz}$ dark count rate. Classical latencies are $\ell_{ij}{=}\frac{L_{ij}}{c_0}{+}\tau_{\rm proc}$ with processing time $\tau_{\rm proc}{=}10\,\mu\text{s}$.

\textit{Memory:}  Each node has an array of 6 quantum memories, each with an efficiency of $\eta_{\rm mem}=0.69$ and a frequency of $50$ kHz. Suppose decoherence is mainly Markovian dephasing with different rates: ${\gamma_0 = 5, \gamma_1=\gamma_2 = 10}$. The success probability of swapping is $q=0.5$. We use $T_{\rm swp} = 2\max(\ell_{01},\ell_{12})$. Node 1 can assign its 6 memories to the links in different ways. We assume \textit{memory multiplexing} for parallel or time-division multiplexed generation, yielding state-dependent arrival rates $\lambda_{i|N_i} = (K_i-N_i)\lambda_i^*$ with $\lambda_0^* = 76.3$ and $\lambda_1^* = 244.5$.

\subsection{Results and Analysis}
\subsubsection{Model predictions}
Fig.~\ref{fig_multi_memory} compares the rate and fidelity predicted by our models against simulation results as we vary the strict fidelity $F_{\rm req}$ for various memory allocations under three matching policies. 
First, in all cases, our models closely match the simulated throughput and fidelity. Besides the BDP, we also implemented its full 4D-CTMC model, validating the approximation used for BDP. 
Second, multiplexing boosts both throughput and achievable fidelity. 
With allocation $(1,5)$, near-maximum throughput is retained until $F_{\rm req} \approx 0.89$ because link 0 is the dominant bottleneck: $\mathcal{Q}_1$ is rarely empty while link 0 generates much more slowly. Reallocating memories from $\mathcal{Q}_1$ to $\mathcal{Q}_0$ relieves the bottleneck, increases the matching rate and throughput, and shortens the wait time at $\mathcal{Q}_1$ to improve fidelity. This is true for all policies. The model captures this shift, accounting for both resource allocation and multiplexing, and easily identifies the optimal allocation $(4,2)$.


\subsubsection{Policy comparisons}
For each memory allocation, all matching policies achieve similar rates, while the distributions/spread of achieved fidelity differ, especially at low $F_{\rm req}$, except for the $(4,2)$ allocation. As expected, LIFO yields the highest mean fidelity, RAND is intermediate, and FIFO is the lowest. The inter-policy gap is most pronounced for $(1,5)$ where link 0 is much slower and $\mathcal{Q}_{1}$ builds up. Thus, FIFO sees larger wait times as it picks the oldest queued entanglement, while LIFO uses the freshest queued entanglement. Allocating more memories to the bottleneck link reduces wait times at the other link, which improves fidelity. This closes the policy gap, with $(4,2)$ being the most balanced. The spread also shrinks as $F_{\rm req}$ increases as MHTs decrease. We note a small systematic bias in the mean fidelity predictions due to the loss of age information in the tagged-pair CTMC. LIFO waiting age is underestimated because pairs that arrive after tagged pair can expire early under the memoryless exponential approximation, allowing the tagged pair to be matched sooner than  possible. On the other hand, FIFO age is overestimated. RAND remains unbiased as it randomly selects a stored pair in a memoryless fashion.

\subsubsection{Probabilistic fidelity guarantee}
Fig.~\ref{fig_F_epsilon} shows the rate and fidelity for $(4,2)$ memory allocation 
as we vary fidelity violation margin $\varepsilon$ with $F_{\rm desired}=0.88$. As expected, as $\varepsilon$ increases, the rate goes up. But, this increase comes at the price of lower goodput, which is the rate of entanglements satisfying $f_{\rm e2e} \geq F_{\rm desired}$. For instance, when $\varepsilon = 0.1$, the rate is approximately 124 (compared to 116 for $\varepsilon < 0.03$), which is 7\% improvement. However, the goodput is 113, which is just above 0.9$\times$124, illustrating an expected trade-off. 

\begin{figure}[t]
    \centering
    \includegraphics[width=0.88\linewidth]{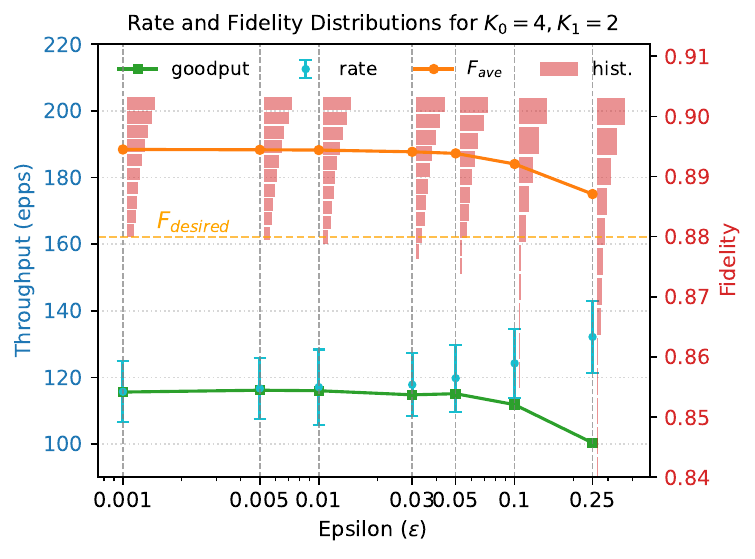}
    \caption{\small Rate and fidelity when fixing $F_{\rm min}=0.8$ but relaxing fidelity violation margin $\varepsilon$ for $F_{\rm desired}=0.88$ under LIFO policy.}
    \label{fig_F_epsilon}
\end{figure}

\section{Concluding Remarks}
This paper developed modeling and optimization tools to maximize entanglement swapping throughput under strict, probabilistic, and average fidelity constraints, while accounting for heterogeneous quantum links with varying decoherence rates, capacities, and communication latencies. By treating the two-link swapping process as a fundamental building block, this work provides a scalable basis for future research into complex architectures. We will extend this analysis to multi-repeater paths and arbitrary network topologies, integrating dynamic routing, memory management, and entanglement purification to support full-scale quantum internet infrastructure.

\appendices
\section{Numerical Optimization for Multi-Memory}
\label{app:Multi-Memory}
The case with strict fidelity admits a simple solution~\eqref{eq_W_threshold}. The problem with probabilistic fidelity is reformulated in~\eqref{prob_maxRate_chanceFidelity}. Let us now extend the mean-fidelity design (Sec.~\ref{subsec_Expected_F}) to multi-memory queues. We also use a general dephasing
\begin{equation}
w_i(t)=e^{-(\gamma_i t)^{\zeta}},\qquad \zeta \ge 1,
\label{eq:gen-deph}
\end{equation}
which recovers Markovian dephasing $\zeta{=}1$, Gaussian dephasing $\zeta{=}2$, and models intermediate non-Markovian environments for $\zeta\in(1,2)$. Note that the throughput model and the
tagged-pair wait-time law are \emph{independent of $\zeta$}; the exponent
$\zeta$ enters only through the per-queue fidelity decay. 

\subsection{Fidelity of a Matched Pair}
Let $\tau_{01}$ and $\tau_{12}$ denote the successful attempt durations, and define $\tau_0:=T_{\mathrm{swp}}+\tau_{01}$ and $\tau_1:=T_{\mathrm{swp}}+\tau_{12}$
denote the fixed heralding-plus-swap ages. When a swap occurs at $\mathcal{Q}_i$ and
the \emph{waiting} pair has waited $\delta$ (the opposite pair being fresh), 
the delivered fidelity 
\begin{equation}
f_i(\delta)=c_1+\tfrac{v_{02}}{2}\,w^{\rm B}_{01}w^{\rm B}_{12}\,e^{-\Psi_i(\delta)} .
\label{eq:fi-general}
\end{equation}
where $c_1=\tfrac{1+v_{02}}{4}$, $v_{02}$ is the depolarizing factor, and the four memory qubits contribute
\begin{align*}
\Psi_0(\delta)
&=[\gamma_0(\tau_0 + \delta + \ell_{01})]^{\zeta}
+[\gamma_1(\tau_0 + \delta)]^{\zeta}
+\chi_1\\
\Psi_1(\delta)
&=[\gamma_2(\tau_1 + \delta + \ell_{12})]^{\zeta}
+[\gamma_1(\tau_1 + \delta)]^{\zeta}
+\chi_0
\end{align*}
with $\chi_0=\gamma_1^{\zeta} \tau_0^{\zeta}+ \gamma_0^{\zeta} (\tau_0+\ell_{01})^{\zeta}$ and $\chi_1=\gamma_1^{\zeta} \tau_1^{\zeta}+ \gamma_2^{\zeta} (\tau_1+\ell_{12})^{\zeta}$. 
Each $\Psi_i$ is strictly increasing and, for $\zeta\ge1$, convex in $x$. Hence, 
$f_i$ is strictly decreasing: a longer wait always lowers fidelity. 
Two special cases $\zeta=1,2$ can be simplified, but for general $\zeta\in(1,2)$ no such reduction exists.

\subsection{Expected E2E Fidelity}
Let $(\nu_i^{\rho},H_i^{\rho},b_i^{\rho})$ be the tagged-pair
initial distribution, subgenerator, and matched-absorption vector under
matching policy $\rho\in\{\textsc{fifo},\textsc{lifo},\textsc{rand}\}$, with density
$\xi_i^{\rho}(x)=\nu_i^{\rho}e^{H_i^{\rho}x}b_i^{\rho}$ and match
probability $C_i^{\rho}(W_i)=1-\nu_i^{\rho}e^{H_i^{\rho}W_i}\mathbf 1$.
Using the mixture wait time distribution, the expected E2E fidelity is given by 
\begin{align}
\mathbb{E}[f_{\mathrm{e2e}}]
&=c_1+\frac{1}{2}\,v_{02}w^{B}_{01}w^{B}_{12}
 \sum_{i=0,1}p_{i|m}\,G_i(W_i),
\label{eq:Ef-general}\\
G_i(W_i)&= \frac{1}{C_i^{\rho}(W_i)}\int_0^{W_i}e^{-\Psi_i(x)}\,
\xi_i^{\rho}(x)\,\mathrm{d}x .
\label{eq:Gi}
\end{align}
Thus, suppressing superscript $\rho$, we have the following. 
\begin{proposition}
If $\zeta=1$, expected fidelity and wait time are
\begin{align}
\mathbb{E}[f_{\mathrm{e2e}}]
&= c_1 + c_2\big(p_{0|m}\Lambda_0(\gamma_{01})
                 + p_{1|m}\Lambda_1(\gamma_{12})\big),
\label{eq:app-Ef}\\
\mathbb{E}[Z]
&= \frac{\gamma_{01}\lambda_{1|0}E_0\,\mu_0
        + \gamma_{12}\lambda_{0|0}E_1\,\mu_1}
       {\lambda_{1|0}E_0+\lambda_{0|0}E_1}.
\label{eq:app-EZ}
\end{align}
where  $\Lambda_i(s) = \mathbb{E}\!\left[e^{-s\delta_i}\,\middle|\,\mathcal{Q}_i\right]$ and $\mu_i
= \mathbb{E}\!\left[\delta_i\,\middle|\,\mathcal{Q}_i\right]$, given by
\begin{align}
\Lambda_i(s)
&\!=\! {\nu_i\,(H_i{-}sI)^{-1} (e^{(H_i-sI)W_i}{-}I)b_i}/
       {C_i(W_i)},
\label{eq:app-Lambda}\\
\mu_i
&\!=\! {\big( \nu_i H_i^{-1}\!(e^{H_iW_i}{-}I )\mathbf{1}
        - W_i\,\nu_i e^{H_iW_i}\mathbf{1}\big)}/{C_i(W_i)}
\label{eq:app-mu}
\end{align}
\end{proposition}
\begin{proof}
The transient generator $H_i$ has eigenvalues with negative real part,
so $(H_i-sI)$ is invertible for $s\ge 0$ and
$$\int_0^{W}e^{-sx}e^{Hx}\,\mathrm{d}x=(H-sI)^{-1}(e^{(H-sI)W}-I)$$ 
giving
\eqref{eq:app-Lambda}. For the mean, integration by parts yields
$$\int_0^{W}x\,e^{Hx}\,\mathrm{d}x = W H^{-1}e^{HW}-H^{-2}(e^{HW}-I)$$ 
multiplying
by $b_i=-H_i\mathbf{1}$ and using $\nu_i\mathbf{1}=1$ gives
\eqref{eq:app-mu}.
\end{proof}
Note that all quantities depend on $W=(W_0,W_1)$ through $E_i(W_i)$ (via $\theta_i=1/W_i$ in
\eqref{eq_pi_0}) and through $H_i,\nu_i$. The following is a useful property.

\begin{lemma}\label{lem:bind}
For every $\zeta>0$ and policy $\rho$, $\mathbb{E}[f_{\mathrm{e2e}}]$ is
strictly decreasing in each $W_i$, while $R_{\rm m}$ is nondecreasing in each
$W_i$. Hence at any optimum of
\begin{equation}
\max \ \{R_{\rm m}(W) : \mathbb{E}[f_{\mathrm{e2e}}](W)\ge F_{\mathrm{mean}} \}
\label{eq:P-general}
\end{equation}
the fidelity constraint is active.
\end{lemma}
\begin{IEEEproof}
The monotonicity of $R_{\rm m}$ comes from Theorem~\ref{thm_rate_multi_mem}. For the expected fidelity, note that increasing $W_0$ raises the truncation point of the sub-density
$\xi_0^{\rho}$, so the conditional matched age $\delta_0$ is
first-order stochastically larger; as $e^{-\Psi_0}$ is decreasing,
$G_0(W_0)$ strictly decreases, while $G_1$ is unchanged. Thus, if the constraint were slack at a candidate point,
increasing the coordinate with positive throughput sensitivity would stay
feasible and raise $R_{\rm m}$, contradicting optimality.
\end{IEEEproof}

\subsection{Numerical Solvers} 
Below are two practical methods that we can employ. 
\paragraph{Nonlinear Programming}
Optimize $W=(W_0,W_1)$ directly in~\eqref{prob_maxRate_chanceFidelity} and \eqref{eq:P-general} using nonlinear methods such as SLSQP. Both $R_{\rm m}$ (closed form), $\Phi$ in \eqref{eq:phi_exact} and $\mathbb{E}[f_{\mathrm{e2e}}]$ in \eqref{eq:Ef-general} are smooth; gradients are obtained by the Fr\'echet derivative of $e^{H_iW_i}$ or central finite differences on the two variables at a numerically negligible cost. If the solver stops a fraction outside its constraint tolerance, we project back onto feasibility, e.g., along the ray $W\mapsto sW$ (fidelity increases as $s$ decreases by Lemma~\ref{lem:bind}), 
guaranteeing $\Phi(W)\le \varepsilon$ and $\mathbb{E}[f_{\mathrm{e2e}}]\ge F_{\mathrm{mean}}$. 
This is fast as the problem is 2D. 

\paragraph{Boundary root-find and line search}
Lemma~\ref{lem:bind} indicates the optimum on the curve
$$\mathbb{E}[f_{\mathrm{e2e}}](W_0,W_1)=F_{\mathrm{mean}}.$$ Since
$\mathbb{E}[f_{\mathrm{e2e}}](\cdot,W_1)$ is strictly decreasing in
$W_0$, for each fixed $W_1$ the equation has at most one root:
\begin{enumerate}
\item sweep $W_1\in[\,W^{\min},W_1^{\max}\,]$;
\item if $\mathbb{E}[f_{\mathrm{e2e}}](W^{\min},W_1)<F_{\mathrm{mean}}$,
      skip (infeasible for every $W_0$); if
      $\mathbb{E}[f_{\mathrm{e2e}}](W_0^{\max},W_1)\ge F_{\mathrm{mean}}$,
      set $W_0=W_0^{\max}$; otherwise solve
      $\mathbb{E}[f_{\mathrm{e2e}}](W_0,W_1)=F_{\mathrm{mean}}$ for $W_0$
      by a bracketed scalar root-finding method;
\item evaluate $R_{\rm m}(W_0,W_1)$ and keep the maximizer.
\end{enumerate}
This returns the global optimum of~\eqref{eq:P-general} in 2D for every $\zeta$ and policy. The main cost is the evaluation of $\mathbb{E}[f_{\mathrm{e2e}}]$. The same approach applies to probabilistic fidelity constrained problem in~\eqref{prob_maxRate_chanceFidelity}.

\end{document}